\documentclass[11pt]{article}
\usepackage{blois2002,psfig}
\def\Journal#1#2#3#4{{#1} {\bf #2}, #3 (#4)}

\def\NCA{\em Nuovo Cimento}

\def\NPB{{\em Nucl. Phys.} B}
\def\PLB{{\em Phys. Lett.}  B}
\def\PRL{\em Phys. Rev. Lett.}
\def\PRD{{\em Phys. Rev.} D}

\def\APJ{\em Astrophys. J.}
\def\PRS{\em Proc. Royal Soc}

\def\be{\begin{equation}}
\def\ee{\end{equation}}
\def\bea{\begin{eqnarray}}
\def\eea{\end{eqnarray}}
\def\epr{e-print astro-ph/}
\def\etal{{\it et al.\ }}
\def\ie{{\it i.e.,}}

\def\gray{$\gamma$-ray\ }
\def\grays{$\gamma$-rays\ }

\begin{document}
\vspace*{4cm}
\title{THE MATTER-ANTIMATTER ASYMMETRY OF THE UNIVERSE}

\author{ F.W. STECKER }

\address{Laboratory for High Energy Astrophysics, NASA Goddard Space 
Flight Center, \\ Greenbelt, MD 20771 USA}

\maketitle\abstracts{I will give here an overview of the present observational
and theoretical situation regarding the question of the matter-antimatter
asymmetry of the universe and the related question of the existence of 
antimatter on a cosmological scale. I will also give a simple discussion of 
the role of $CP$ violation in this subject.}

\section{Introduction}

One of the most fundamental questions in cosmology is that of the role of 
antimatter in the universe. This question, which is intimately connected to the
question of the nature of $CP$ violation at high energies, is the important
subject of this conference. It is a question for theorists, but ultimately
as in all scientific endevours, a question which must be answered empirically 
if possible. 

The discovery of the Dirac equation \cite{di28} placed antimatter on an equal
footing with matter in physics and opened up speculation as to whether there
is an overall balance between the amount of matter and the amount of 
antimatter in the universe. The hot big bang model of the universe added a
new aspect to this question. It became apparent that in a hot early epoch of
the big bang there would exist a fully mixed dense state of matter and 
antimatter in the form of leptonic and baryonic pairs in thermal equilibrium
with radiation. As the universe expanded and cooled this situation would 
result in an almost complete annihilation of both matter and antimatter.

The amount of matter and antimatter expected to be left over in an expanding
universe can be calculated from the proton-antiproton annihilation cross 
section. Antinucleons ``freeze out'' of thermal equilibrium when the 
annihilation rate becomes smaller than the expansion rate of the universe. 
This would have occurred when the temperature of the
universe dropped below $\sim 20$ MeV. The predicted freeze out density of
both matter and antimatter is only about $4\times 10^{-11}$ of the closure 
density of the universe ({\it i.e.}, $\Omega_{baryon} = 4\times 10^{-11}$) 
\cite{ch66}.

\begin{figure}
\centerline{\psfig{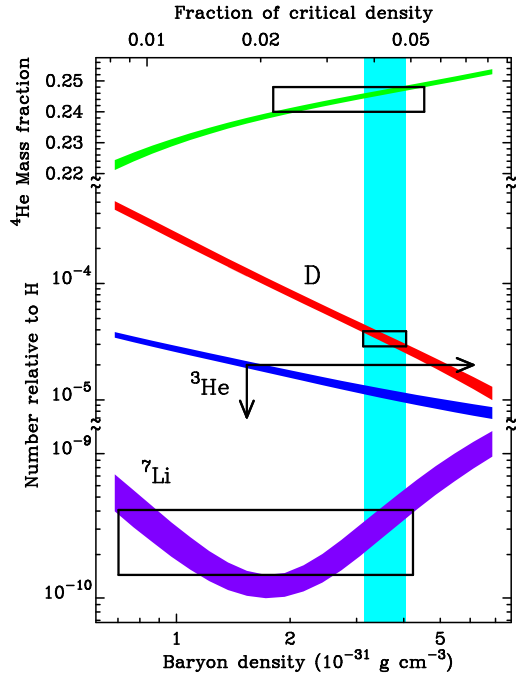}}
\caption{Predicted abundances of light nuclides from big-bang nucleosynthesis.
\protect\cite{bu99} \label{bbn}}
\end{figure}

On the other hand, big-bang nucleosynthesis calculations \cite{ol02}
and studies of the anisotropy of the 2.7 K cosmic background radiation 
\cite{si02} have indicated that baryonic matter makes up about 4\% of the 
closure density of the universe, ({\it i.e.}, 
$\Omega_{baryon} \simeq 4 \times 10^{-2}$) as
shown in Figure \ref{bbn}.  
Thus, there is a nine order of magnitude difference between the simple
big-bang prediction and the reality of the amount of baryonic matter which
is found in the universe and which makes up the visible matter in galaxies
as well as the matter in you and me. Clearly, there is something missing.
It was elegantly shown by Sakharov \cite{sa66} that what is missing
is the breaking of symmetries. In order to make an omelet you have to break
some eggs; in order to make a universe you have to break some symmetries.
It is in this context that the question of the nature of the violation of
$CP$ symmetry arises.

\section{The Sakharov Conditions (and Beyond).}

Sakharov showed that three conditions are necessary in order to create
the appropriately significant concentration of baryons in the early universe.
They are:

\vspace{0.6cm}

\noindent$\bullet$ {\bf Violation of Baryon Number, $B$}

\vspace{0.5cm}

\noindent$\bullet$ {\bf Violation of $C$ and $CP$}

\vspace{0.5cm} 

\noindent$\bullet$ {\bf Conditions in which Thermodynamic Equilibrium does not
Hold}

\vspace{0.6cm}

The first condition is satisfied in grand unified theories (GUTs) in which
strong and electroweak interactions are unified and quarks and leptons are
placed in the same multiplet representations. It is also satisfied in
electroweak theory through the sphaleron mechanism (see next section).

The second condition involves the nature of $CP$ violation ($CPV$). 
We know that $CP$ is violated at low energies in $K^{0}$ mixing and decay
\cite{ch64} and in $B^{0}$ mixing \cite{au02}. These processes are  
a major subject of this meeting. The creation of matter and antimatter 
asymmetries in the universe (baryogenesis) involves 
the nature of $CP$ violation at high energies. 
As will be discussed in the next section, the standard $SU(3) 
\times SU(2)_{L} \times U(1)_{Y} $ model cannot account for baryogenesis at 
high energies. This implies
that there is no obvious relationship between the low energy $CPV$ 
which has been observed in the laboratory and which can be described by the CKM
(Cabibbo-Kobayashi-Maskawa) \cite{ca63} matrix and that which 
must account for baryogenesis. 

The third condition of non-equilibrium can be supplied at the GUT scale by 
the expansion of the universe. Owing to this expansion, below the GUT 
temperature, $CP$ violating decays of leptoquarks or GUT-Higgs bosons
cannot be balanced by their inverse reactions. At the electroweak scale,
things are different (see next section).

The three Sakharov conditions are part of the recipe for making our universe
omelet. However, the nature of the $CPV$ is also important. If it is
spontaneous $CPV$ \cite{br79}, followed by a period of moderate
inflation \cite{sa81}, one can generate astronomically large domains of $CP$
violation of opposite sign. In principle, subsequent baryogenesis can lead
to separate regions containing matter galaxies and antimatter galaxies 
respectively (see section 4).

\section{Baryosynthesis}
\subsection{Different Baryogenesis Scenarios}

As we have already mentioned, baryon number is naturally violated in grand
unified theories. It is also violated in the 
$SU(3) \times SU(2)_{L} \times U(1)_{Y} $ 
standard model (SM) because in this model gauge invariant chiral
currents are not conserved owing to the Adler-Bell-Jackiw anomaly. \cite{ad69}
As a result of this anomaly, 
together with the fact that weak gauge bosons couple 
only to left handed quarks and leptons, it can be shown that the SM conserves
$(B-L)$ but violates $B$ and $L$ separately. At electroweak unification
scale temperatures $T_{EW} \sim$ 100 GeV, SM $B$ and $L$ violation can occur 
freely through ``sphaleron'' transitions between topologically distinct
vacuum states with neighboring winding numbers. \cite{ku85} This transition 
process is suppressed at temperatures much less than $T_{EW}$ by an 
exponential barrier penetration factor. 

However, at a temperature $T_{EW} \sim 100$ GeV,
the expansion of the universe is too slow relative to
the weak and electromagnetic interaction rates for Sakharov's 
out-of-equilibrium condition to hold. The weak interaction rate, 
$\Gamma_{weak}$ is proportional
to $\sigma_{weak} \simeq G_{F}^2T^2$ times the particle density 
$n$, where $G_{F} \simeq 10^{-5}$ 
GeV$^{-2}$. Here, we adopt natural units ($h/2\pi = c = k = 1$) with the 
temperature in  GeV. With these units, within an order of magnitude, 
the particle density, $n \sim T^3$ and the rate of expansion of the universe 
$H \sim T^2/M_{Planck}$ in the radiation dominated era. The ratio,

$$ {\Gamma_{weak}\over{H}} ~ \sim ~ G_{F}^2M_{Planck}T_{EW}^3  ~ \sim ~ 10^{15}. $$

\noindent For electromagnetic interactions, this ratio is 
even larger. Thus, the expansion of the universe is much too slow to break 
thermal equilibrium.

For effective baryogenesis to occur, Sakharov's condition of thermal 
non-equilibrium must be adequately met by another means. This requirement
is satisfied if the Higgs fields undergo a strongly first
order phase transition when the electroweak symmetry is broken. \cite{ku85}
In order for such a phase transition to occur, lattice simulations 
have shown that the
required SM Higgs mass must be less than $\sim$ 72 GeV. \cite{ru98} However,
lower limits on the mass of the electroweak Higgs boson obtained at LEP
indicate that $m_{H} \ge 110$ GeV. \cite{ab01} This precludes the required
phase transition in the case of the SM. However, 
it has been proposed that extensions of the SM with extra Higgs fields
may work. \cite{be02} 

In any case, it is doubtful whether the $CPV$ described
by the CKM matrix plays a role at high temperatures in the early universe
or is related to the $CPV$ needed for baryogenesis. 
It has been shown that the $CPV$ provided by the CKM matrix cannot produce a 
large enough baryon asymmetry because, owing to GIM suppression \cite{gl70}, 
its contribution to baryon number violation
only arises at the three loop level. \cite{ba79} 
It is more likely that $CPV$ involving GUT scale mechanisms comes into play. 
These mechanisms can also involve the Higgs sector.

\subsection{Example: The Weinberg Scenario}

A simple scenario for baryon production from superheavy particle decay which
can serve as an illustration of baryogenesis was given by Weinberg \cite{we79}.
Weinberg's GUT-inspired X particles decay {\it via} two channels with
baryon numbers $B_{1}$ and $B_{2}$ and branching ratios $r$ and $1-r$. 
The antiparticles decay with baryon numbers $-B_{1}$ and $-B_{2}$ with the same
total rate, but with different branching ratios $\bar{r}$ and $1-\bar{r}$.  
The mean net baryon number produced is then

$$\Delta B = (1/2)(r - \bar{r})(B_{1} - B_{2}]. $$

The baryon to photon ratio produced is estimated by Weinberg to be two to
three orders of magnitude below $\Delta B$. This factor is arrived at by
noting that all of the particle densities started out equal in thermal 
equilibrium and taking account of the fact that
it is the surviving baryon to entropy ratio which is conserved during
the subsequent expansion of the universe and that the entropy at the time
of baryogenesis, \ie the GUT era, was larger by roughly $10^2 - 10^3$, counting
the additional degrees of freedom supplied by the particles of mass
$m << M_{GUT}$.

As an even simpler example than the above one, consider a GUT leptoquark boson
with the decay modes $X \rightarrow ql$ with branching ratio $r$ and
$X \rightarrow \bar{q}\bar{q}$ with branching ratio $(1-r)$. Then $\Delta B
= (1/2)(r - \bar{r}).$ If $CP$ is not violated, then $r = \bar{r}$ and
$\Delta B = 0$. Thus, we require $CP$ violation. But, also note that
the {\it sign} of $\Delta B$ depends on the {\it sign} of $CP$ violation. 
This will be a critical point in the next section.  

It has been pointed out that since the spaleron mechanism conserves $(B - L)$
(see above), any baryosynthesis involving a GUT gauge group containing a
$U_{B-L}$ symmetry as a subgroup which is unbroken above the GUT scale will 
be washed 
out at the electroweak level by sphaleron interactions. \cite{ku85} 
Recently, models have been proposed where lepton number is violated at the
GUT scale and then, as the universe cools, baryon number is generated 
through the sphaleron mechanism. \cite{fu86}  
However, even more recently, Weinberg-type processes involving GUT particle
decay have been resurrected 
in scenarios involving Majorana neutrinos. \cite{fu02} 

Other discussions of baryogenesis mechanisms are given by Dolgov and
Berezhiani in these proceedings.

\section{A Locally Asymmetric Domain Cosmology}

If $CPV$ is predetermined, then only matter will remain in the 
present universe. We can refer to this case as
a ``global'' matter-antimatter asymmetry. If, on the other hand, $CPV$ is the 
result of spontaneous symmetry breaking, domains of positive and negative 
$CPV$ may result \cite{br79}. In the case of spontaneous $CPV$, the Lagrangian 
is explicitly $CP$ invariant, but at the symmetry breaking phase transition 
a $CP$ invariant high temperature vacuum state undergoes a transition to a 
state where the vacuum solutions break $CP$ either way. \cite{br79} \cite{se80}
\cite{st81} This mechanism may be compared to the spontaneous formation of 
ferromagnetic domains when a piece of unmagnetized iron cools below the 
critical 
temperature in the absence of a magnetic field. Although there is no preferred
direction of magnetization, individual domains acquire random local directions
of magnetization.

If the $CP$ domain structure is stretched to astronomical size by a 
subsequent period of moderate inflation \cite{sa81},
then, following baryogenesis, baryons may survive as galaxies in 
some regions of the universe 
and antibaryons may survive as antigalaxies in other regions. In this
case, we have a ``local'' matter-antimatter asymmetry instead of a
global one. We will refer to this possibility as a ``locally asymmetric domain 
cosmology (LADC).'' Following baryogenesis, the walls of the initially
$CP$ symmetric vacuum between the positive and negative $CP$ domains must 
vanish because they are quite massive and could eventually 
dominate the evolution of the universe, in conflict with observations.
\cite{ze74} Various mechanisms have been proposed to accomplish this.
\cite{vi81} \cite{ku81} \cite{mo84}
The imprint of the $CP$ domains remains as ``fossil'' baryon and antibaryon 
``domains''.\footnote{A cosmological model where antimatter plays a minor role
has also been considered more recently. \cite{kh00}}

Unfortunately, we cannot aim the Hubble Space Telescope at distant galaxies
in order to find antimatter galaxies. 
Antimatter galaxies will look exactly the same as matter galaxies.
This is because the photon is its own antiparticle. 
However, we can look for other clues. Searches have been made for antimatter
in the cosmic radiation and for the indirect traces of cosmic matter-antimatter
annihilation in the extragalactic $\gamma$-ray background radiation. The
results of these searches will be discussed in the next two sections.

\section{Antimatter in the Cosmic Radiation}

\begin{figure}
\centerline{\psfig{figure=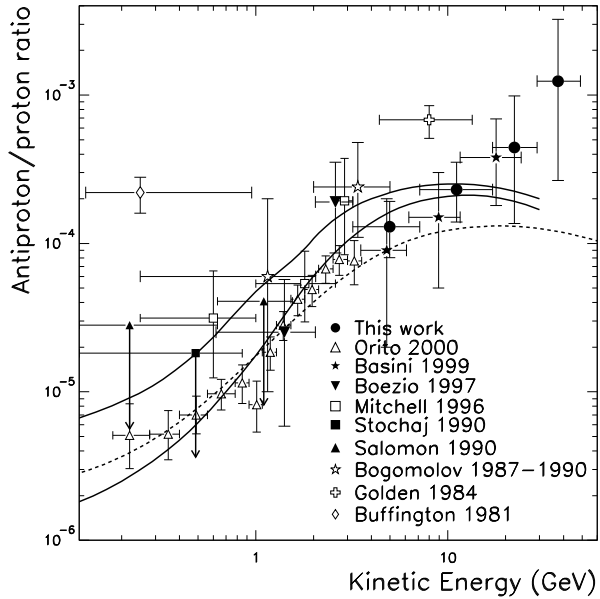,height=3.5in}}
\caption{Measured cosmic ray $\bar{p}$/$p$ flux ratios 
and limits as a function of energy. \protect\cite{bo01}
\label{pbar}}
\end{figure}

Many measurements have been made of antiprotons in the cosmic radiation.
A recent compendium of measurments of the $\bar{p}$/$p$ flux ratios as a
function of energy is given in Figure \ref{pbar}.
This figure also shows curves of the predicted flux ratios obtained
by calculating the production of secondary antiprotons from galactic cosmic ray
interactions with interstellar gas nuclei. As the figure indicates, the present
measurements up to $\sim 40$ GeV are consistent with secondary production.
(See also the paper of Coutu in these proceedings. \cite{bea01})

\begin{figure}
\centerline{\psfig{figure=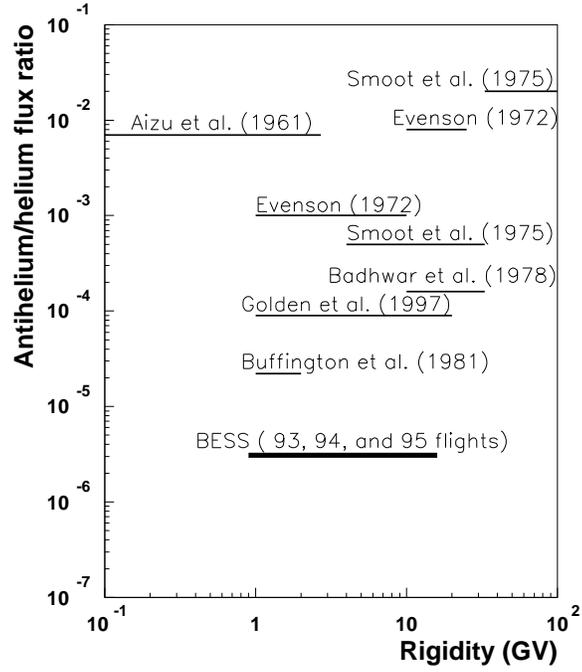,height=3.5in}}
\caption{Observational limits on the flux ratio of $\bar{\alpha}$/$\alpha$ as 
a function of energy. \protect\cite{sa98}
\label{abar}}
\end{figure}

\begin{figure}
\centerline{\psfig{figure=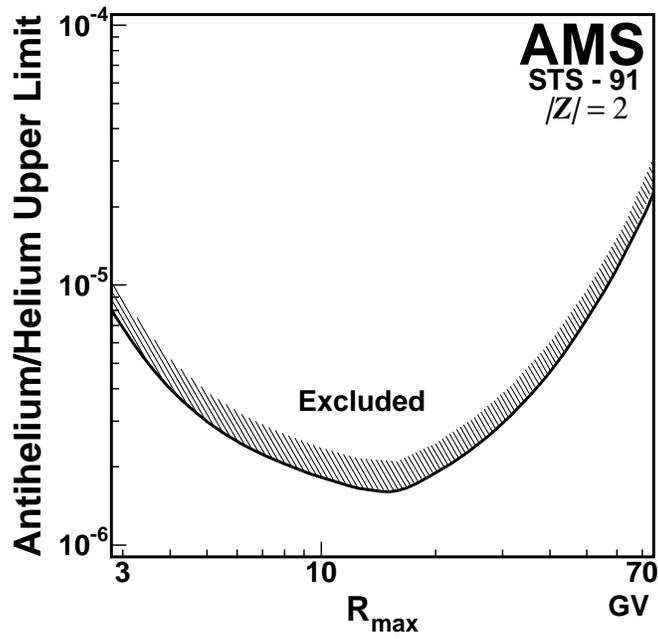,height=3.5in}}
\caption{Recent upper limits on the flux ratio of $\bar{\alpha}$/$\alpha$ as a 
function of energy obtained by the Alpha Magnetic Spectrometer (AMS) detector.
\protect\cite{al99}
\label{ams}}
\end{figure}

There have been no antihelium nuclei detected in the cosmic radiation. Recent
limits on the antihelium-helium ratio are shown in Figures \ref{abar} and
\ref{ams}.

Thus, there is no evidence for extragalactic antimatter in the low energy
cosmic rays. This is, of course, direct evidence against the existence of 
significant antimatter in our own galaxy since galactic \gray observations
indicate that cosmic rays diffuse throughout our galaxy. \cite{st77} Studies
of secondary nuclides produced by cosmic ray interactions with interstellar
gas also indicate diffusion of comsic rays throughout the galaxy. \cite{str98}

As indicated in Figures \ref{abar} 
and \ref{ams}, the most stringent limits on cosmic ray $\bar{\alpha}$'s are 
at very low energy. The limits 
get worse as the energy goes up because of the decrease in detector 
sensitivity owing to the smaller bending of high energy particles (and 
antiparticles) in the magnetic field of the detectors. 

Stecker and Wolfendale \cite{st84} have shown 
that the ratio of extragalactic cosmic rays to galactic cosmic rays should 
increase with energy owing to the fact that the escape rate of galactic cosmic
rays from the galaxy increases with energy. In addition,  
there is a significant question as to whether extragalactic cosmic
rays can enter the galaxy owing to the presence of a galactic wind.
Ahlen \etal \cite{ah82} have estimated that the galactic wind could
reduce the low energy component of the extragalactic cosmic ray 
flux by more than an order of magnitude. Thus, a better test for extragalactic
antimatter would come with the measurement of cosmic rays at much higher
energies. There is also the question as to whether cosmic rays can diffuse and
propagate to Earth from the distances of tens of Mpc required by the LADC
models (see next section).

\section{The MeV Gamma Ray Background Test}

Perhaps the most significant and potentially observable consequence
of LADC is the prediction of a $\gamma$-ray background from the
annihilation of matter and antimatter taking place at the boundaries
between matter and antimatter regions. In fact, the possiblity that 
this effect could explain
the multi-MeV background observations was the original motivation
for the author's work on this topic. \cite{st71}

Another prediction of the LADC is that the \gray background radiation
from boundary annihilations would not really be isotropic. There would
be a structure of \gray ``ridges'' produced and these ridges would be
more pronounced at energies near 100 MeV than at MeV energies.
\cite{ga90} 

The source of the MeV-range background radiation is still a mystery.
It has been suggested that redshifted line emission from extragalactic
supernove can
explain part of the flux \cite{wa99}. However, such radiation is limited to
the energy range below 3.5 MeV and therefore cannot account for all of the
flux. Another suggestion has been the superposition of MeV emission tails
from active galaxies \cite{st01}. In  this regard, it should be noted that  
previously flown \gray 
telescopes were not sensitive enough to determine if such non-thermal
multi-MeV \gray tails are produced by individual active galaxies. The 
theoretical situation regarding the origin of the multi-MeV \gray background 
is therefore unclear at this point.

\begin{figure}
\centerline{\psfig{figure=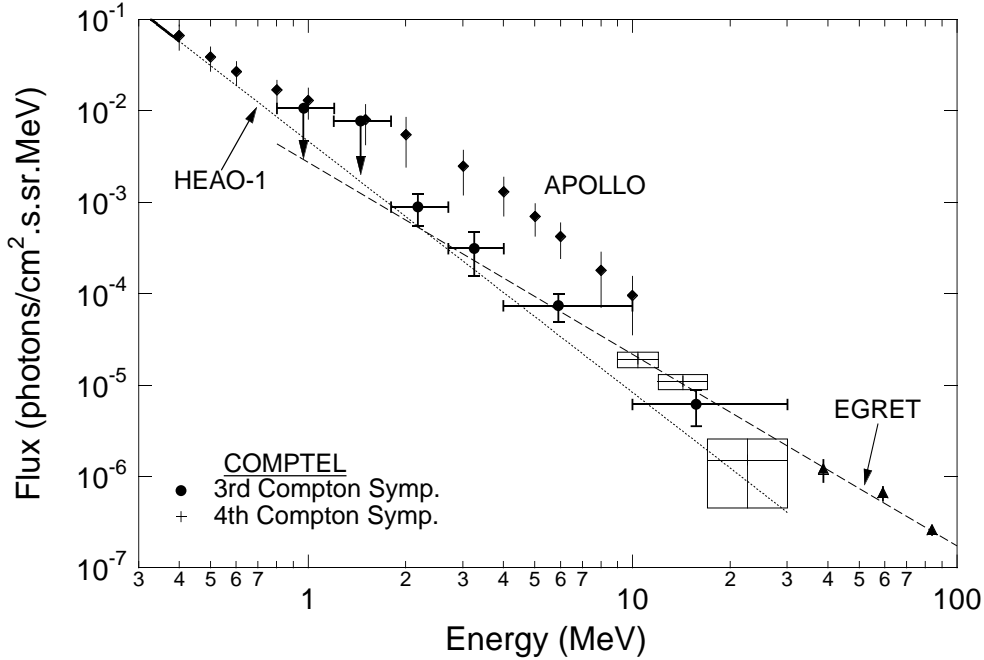,height=3.5in}}
\caption{A compliation of diffuse $\gamma$-ray background measurements as a 
function of energy. \protect\cite{sr97}
\label{grb}}
\end{figure}

The observational situation is summarized in Figure \ref{grb}. The two
results from Apollo 16 and 17 and Comptel both involved the difficult 
determination of a subtraction of the flux of \grays produced in the 
detector 
and surrounding material which was much larger than the true signal itself. 
Of course, in such a case, there is always the danger of oversubtraction
as well as undersubtraction. The result is that the two data sets are, 
in some places, more than an order of magnitude in disagreement. 

Based on preliminary Apollo 15 data, Stecker and Puget \cite{st72} 
estimated the size of the fossil matter and antimatter domains to be of the 
order of 10 Mpc, \ie at least the size of galaxy superclusters. 
Using the method of Stecker \etal \cite{st71}, Cohen 
\etal \cite{co98} estimated a fossil domain size of at least 1 Gpc. Neither of 
these papers took account of the possibility that a magnetic field which 
might be generated parallel to boundary might inhibit the diffusion of 
particles across the boundary and decrease the estimated annihilation rate. 

A new dedicated satellite MeV \gray detector is required to clarify the
observational situation and to determine the flux and spectrum of \grays
in this critical energy range.  The satellite should be light-weight and
contain only the MeV detector and no other experiments in order to
minimize the mass in which cosmic rays can induce intrinsic MeV photon
production.  It should be also flown in a region far from
the Earth's radiation belts, since such radiation induces intrinsic MeV photon 
production within the satellite and detector.

\section{Conclusions}

The fact that simple hot big bang ``freeze out'' calculations predict
a baryon density in the universe which is nine orders of magnitude too
low indicates that $B$, $C$ and $CP$ symmetries must be broken in the
early universe at times corresponding to a temperature greater than 20 MeV,
the simple freeze-out temperature for nucleons and antinucleons. The violation
of these symmetries, especially $CP$, and their consequences for cosmology
are the subject of this meeting. If $CP$ violation is predetermined, than
only matter will remain in the present universe. If, on the other hand,
$CP$ violation ($CPV$) is the result of spontaneous symmetry breaking, domains
of positive and negative $CPV$ may result. If this domain structure is
stretched to astronomical size by a subsequent period of moderate inflation,
then fossil baryons may survive as galaxies in some regions of the universe 
and fossil antibaryons survive as antigalaxies in other regions. We have
referred to this possibility as ``locally asymmetric domain cosmology 
(LADC).'' A longer period of inflation would result in the entire visible
universe being in one domain region.

As of this writing, there is no evidence for large scale extragalactic 
antimatter and, by inference, for LADC. Cosmologically significant
sub-galaxy size antimatter regions are ruled out by their 
potential effect on big-bang nucleosynthesis.
\cite{ku00} Significant antimatter in our own galaxy is ruled 
out by low energy cosmic ray measurements. Although presently unclear,
\gray background measurements indicate that in a LADC cosmology, the size
of the separate regions of matter and antimatter must be at least of galaxy
supercluster extent. 

However, in the present search for cosmological antimatter, 
absence of evidence is not necessarily evidence of 
absence. A dedicated MeV background satellite detector experiment
designed to be as clean from radiation induced intrinsic contamination 
as possible 
would help to clarify the situation. A possible determination of departures 
from isotropy at 20 MeV by the GLAST (Gamma Ray Large Area Telescope) 
satellite, to be launched in 2006, may provide another test. However, this 
test is compromised by the real possibility that the 20 MeV background may be
dominated by unresolved blazars \cite{st96}. Another interesting test would
be to look for departures from isotropy in the cosmic background radiation
caused by the interactions of high energy electrons from the decay of
$\pi^{\pm}$'s produced by annihilation at the boundaries of matter and 
antimatter regions. \cite{ki97} 

\section{A Final Thought}

I stated in the introduction section that the question of the existence
of antimatter in the universe, as with all fundamental physics questions,
in the end must be answered empirically. The discovery of even one ``gold
plated'' antihelium nucleus in the cosmic radiation could change our whole
outlook on this question. I remark that a fish called the coelacanth was
believed to have become extinct 65 million years ago until one was discovered
in the 1930s. The discovery of an antihelium ``helicanth'' would have a
much more profound effect on science.



\end{document}